\documentclass{Interspeech}

\interspeechcameraready

\title{Unveiling Audio Deepfake Origins: A Deep Metric learning And Conformer Network Approach With Ensemble Fusion}

\author[affiliation={1}]{Ajinkya}{Kulkarni}
\author[affiliation={2}]{Sandipana}{Dowerah}
\author[affiliation={2}]{Tanel}{Alumäe}
\author[affiliation={1}]{Mathew}{Magimai.-Doss}

\affiliation{}{Idiap Research Institute}{Switzerland}
\affiliation{}{Tallinn University of Technology}{Estonia}
\email{ajinkya.kulkarni@idiap.ch, sandipana.dowerah@taltech.ee, tanel.alumae@taltech.ee, mathew@idiap.ch}
\keywords{source tracing, audio deepfake, out-of-domain}

\usepackage{comment}
\usepackage{amsmath}  
\usepackage{amssymb}  
\usepackage{mathtools} 
\usepackage{graphicx}
\usepackage{booktabs}
\usepackage{multirow}
\usepackage{tabularx}
\usepackage{textcomp}
\usepackage[utf8]{inputenc}
\usepackage{tipa}
\begin{document}

\maketitle


\begin{abstract}

Audio deepfakes are acquiring an unprecedented level of realism with advanced AI. While current research focuses on discerning real speech from spoofed speech, tracing the source system is equally crucial. This work proposes a novel audio source tracing system combining deep metric multi-class N-pair loss with Real Emphasis and Fake Dispersion framework, a Conformer classification network, and ensemble score-embedding fusion. The N-pair loss improves discriminative ability, while Real Emphasis and Fake Dispersion enhance robustness by focusing on differentiating real and fake speech patterns. The Conformer network captures both global and local dependencies in the audio signal, crucial for source tracing. The proposed ensemble score-embedding fusion shows an optimal trade-off between in-domain and out-of-domain source tracing scenarios. We evaluate our method using Frechet Distance and standard metrics, demonstrating superior performance in source tracing over the baseline system.


\end{abstract}





\section{Introduction}

'\textit{Don't believe everything you hear}', holds true in today's highly digitized world. Technology has made our lives much easier on one hand, and much harder to believe what we hear and see. Highly realistic content generated by sophisticated text-to-speech (TTS) \cite{knntts} and voice conversion systems \cite{knnvcr} has led to the growth of deepfake audios, one of the most intriguing and concerning developments today. Deepfakes are becoming a real threat in scamming people, compromising privacy and social integrity. As these technologies are preceded by the uncanny imitation of 'human speech' with remarkably crafted realistic waveforms, the ability to analyze raw audio data becomes essential in differentiating authentic recordings from manipulated ones. Most studies for audio deepfake detection focus on differentiating between real and fake speech \cite{todisco19_interspeech, kulkarni24_asvspoof, Wang2019ASVspoof2A}. However, it is also important to trace the source system of the generated speech to better detect the manipulated content, especially in out-of-domain attack scenarios, thus, ensuring accountability, and protecting intellectual property.

Deepfake audio source tracing is also crucial in combating manipulative audio content. It aims to identify the system responsible for generating a given utterance. Borrelli et.al. \cite{Borrelli2021SyntheticSD} focused on system identification in ASVspoof 2019 \cite{Wang2019ASVspoof2A} but overlooked categorization based on system attributes. Zhang et.al. \cite{zhang2024distinguishingneuralspeechsynthesis} classified spoofing systems by vocoder type. Zhu et.al. \cite{zhu2022sourcetracingdetectingvoice} attempted multi-attribute classification, including the vocoder, speaker representation, and conversion model. However, distinguishing sources of synthetic audio remains challenging. 

Recent advancements in Audio source tracing methods primarily focus on in-distribution classification \cite{Klein2024SourceTO}. However, as deepfake generation techniques continue to evolve, identifying novel out-of-distribution (OOD) deepfake categories has become increasingly essential. To tackle this challenge, the Audio Deepfake Detection Challenge 2023 Track 3 (ADD2023T3) was introduced \cite{Yi2023ADD2T}. It emphasizes the detection of diverse fake audio types while accounting for unknown generative algorithms during testing. This requires models not only to classify known in-distribution categories but also to distinguish OOD samples effectively. In 2025, the neural audio source tracing task was presented by \cite{nast} established a comprehensive benchmark for evaluating source tracing models. 

State-of-the-art approaches in the ADD2023T3 challenge commonly adopt multi-class classifiers that leverage various features and model architectures. During testing, OOD detection techniques identify unknown classes and classifier scores are fused for evaluation \cite{baseline_refd}. However, these methods present several challenges. First, a single classifier struggles to balance the different feature requirements for the real-vs-fake classification and to distinguish between various fake categories, making training more complex. Second, in such methods, the real class is part of an in-distribution, making it difficult to determine a clear OOD threshold. In ADD2023-T3, OOD samples originate solely from unknown deepfake algorithms, leading to both semantic shifts (real vs. fake) and covariate shifts (distribution differences in fake categories), complicating the detection process. Lastly, while fusion techniques help incorporate different feature representations, they often require extensive fine-tuning of model parameters on the test set, making them inefficient and less adaptable in real-world scenarios.

In this work, we propose a robust approach that integrates novel evaluation metrics, self-supervised learning (SSL) feature extraction techniques, and ensemble fusion strategies. Our method leverages Fr\'echet Distance \cite{FD1,FD2} for OOD-relevant evaluation in audio source separation, offering a deeper understanding of how well-separated sources retain their original characteristics. Unlike traditional metrics, it captures distributional similarities, making it particularly effective in tracing tasks, where subtle variations are critical. We incorporate Conformer \cite{conformer}, which models local and global dependencies, to enhance speaker-specific feature extraction. Additionally, we integrate deep metric multi-class N-Pair Loss \cite{npair} in an adversarial setting to improve the disentanglement of synthetic audio sources, enabling the system to learn embedding latent space clustered with to audio sources while maintaining intra-class consistency. We also implement ensemble fusion using state-space models (MAMBA \cite{mamba}, HYDRA \cite{hydra}), and Conformer architectures, leveraging their complementary strengths for better generalization. By systematically comparing back-end models, we optimize fusion strategies, ensuring adaptability to diverse tracing challenges. Our approach combines Wav2Vec2-XLSR SSL \cite{Baevski2020wav2vec2A} and ensemble fusion to improve accuracy and robustness in tracing synthetic speech to its source. Experimental evaluation demonstrates that fusion of the XLSR-Conformer variants enhances feature representation, while multi-class N-Pair Loss refines the discrimination of audio source embeddings. The ensemble fusion strategy further boosts performance by leveraging diverse model strengths, leading to improved generalization across both in-domain and OOD scenarios. This comprehensive framework effectively addresses the challenges of unknown algorithm variations and dataset shifts, in scaling audio source tracing to real-world applications.

\section{Proposed approach}

In this section, we present the proposed framework for audio source tracing and OOD detection. We begin by providing a brief overview of the classification networks, the feature extraction method, and the Deep Neural Network (DNN) models. Next, we elaborate on the training strategies incorporated in our approach, including the novel deep metric loss criteria within the Real Emphasis and Fake Dispersion framework, as proposed in \cite{baseline_refd}. Finally, we discuss the integration of an OOD detector for inference.

\subsection{Classification Network}



Wav2Vec2-XLSR\footnote{\url{https://huggingface.co/facebook/wav2vec2-xls-r-300m}} \cite{Baevski2020wav2vec2A} along with various back-end DNN architectures have shown state-of-the-art performances in audio deepfake detection \cite{tcm,sls,mambadf}. Therefore, we opted for the Wav2Vec2-XLSR as a feature extraction pipeline in the classification network. During the training phase, we did not freeze the model parameters of Wav2Vec2-XLSR to optimize the self-supervised learned feature representation efficiently into the audio source tracing task. We extracted the last layer of the hidden states as features to classifier back-end DNN of $1024$ dimensions. After the feature extraction step, we provided SSL features to a DNN back-end, where we experimented with three different network architectures: Conformer \cite{conformer}, MAMBA \cite{mamba}, and HYDRA \cite{hydra}. The hidden outputs of these architectures were then passed through an output feedforward layer to predict the probability distribution of classification labels for synthetic audio sources. Conformer is a hybrid architecture that combines convolutional neural networks (CNNs) and Transformers. Its ability to efficiently model both time and frequency relationships makes it particularly useful for classification tasks involving speech and audio processing \cite{tcm,sls,conformersv}. 

MAMBA and HYDRA are state-space models designed to process long-range dependencies in sequential data more efficiently than traditional Transformers \cite{mamba,hydra}. Unlike Transformers, which rely on self-attention with quadratic complexity, MAMBA and HYDRA achieve linear-time complexity, enabling them to handle long-range dependencies with lower computational costs and greater scalability \cite{transformer}. HYDRA is a bi-directional variant of the MAMBA architecture, further enhancing its ability to capture contextual information from past and future sequences \cite{hydra}. 


\subsection{Training stage}

We adopted a two-stage training framework based on the Real Emphasis (RE) and Fake Dispersion (FD) training strategy proposed by Xie et al. \cite{baseline_refd}. In the first stage, training utterances were labelled as real audio with the label $0$ and fake audio with the label $1$, treating the classifier training as a binary classification task using the binary cross-entropy loss criterion. This training phase leverages One-Class (OC) Softmax \cite{ocsoftmax} to adjust the boundary parameters before transitioning to the fake dispersion training stage. Consequently, the RE stage plays a crucial role in mitigating domain gaps and learning meaningful representations of latent space embeddings associated with the fake audio. 




In the FD stage, we employed two loss criteria based on a mixup strategy and a contrastive learning framework. The RegMixup strategy is derived from the principles of Vicinal Risk Minimization \cite{vrm} and incorporates elements of Empirical Risk Minimization \cite{erm}.  In RegMixup \cite{regmix}, new data points are synthesized around the label class distribution, enhancing cross-validation and optimizing soft proxy labels to maximize entropy. This approach facilitates label data distribution, introducing angular bias between labels and out-of-distribution data points, thereby improving model generalization. The loss of Regmixup is defined as below,

\begin{equation}
\mathcal{L}_{RegMixup} = \text{CE} \left( p_{\theta} (\hat{y} \mid x_i), y_i \right) + \eta \, \text{CE} \left( p_{\theta} (\hat{y} \mid \overline{x}_i), \overline{y}_i \right),
\end{equation}

\noindent where \( x_i \in \mathbb{R}^{D} \) and \( y_i \in \{0,1\} \) are output embedding from classification network and audio source labels respectively. For each sample \( x_i \) in a batch, another sample \( x_j \) is randomly chosen from the same batch to create the interpolated representations \( \overline{x}_i \) and \( \overline{y}_i \). The term CE represents the standard cross-entropy loss function. To handle out-of-distribution scenarios, we propose to augment multi-class N-pair loss \cite{npair} as deep metric learning \cite{metriclearning} along with Regmixup, as shown below, 

\begin{equation}
\mathcal{L}_{N-pair} = \beta \cdot \log \left( 1 + \sum_{i=1}^{N-1} \exp \left( x^\top x_i^{-} - x^\top x^{+} \right) \right)
\end{equation}

\noindent where $x$ is embedding from classification network and $\beta$ is weight factor applied to $\mathcal{L}_{N-pair}$. The multi-class N-pair loss enhances the latent representation compared to triplet loss by pushing away multiple negative samples at each backpropagation update step. This results in increasing the intercluster distance from $N - 1$ negative samples and decreases the intracluster distance between positive samples and training examples. In this case, $N$ refers to the number of audio source labels in training, positive samples refer to embeddings from the same audio source labels class and negative samples correspond to embeddings sampled from different audio source labels in the training setup.

\begin{equation}
\mathcal{L}_{FD} = \mathcal{L}_{RegMixup} + \mathcal{L}_{N-pair}
\end{equation}


\subsection{Inference stage}

We utilized Novel Similarity Detection (NSD) \cite{baseline_refd} as an OOD detector to identify the sources of audio deepfakes. During the inference stage, we extract embeddings and classification scores (logits) from the training and test sets. NSD treats all training data as a known category and measures similarity between embeddings from the training, development, and test sets using cosine similarity. Next, confidence scaling is applied to classification similarity scores to smooth the distribution of test set scores. Finally, a threshold is determined using the development set and applied to the test set to classify whether a given test sample originates from a novel deepfake algorithm.


\section{Experimentation}

\subsection{Dataset}

This work utilizes the MLAAD dataset \cite{mlaad} for the audio source tracing task. MLAAD is a multilingual dataset originally designed for audio anti-spoofing. It is intended to assess the out-of-domain generalization of anti-spoofing systems across both new languages and unseen TTS models. MLAAD includes $378$ hours of synthetic speech spanning $38$ languages, generated using $82$ TTS models built on $33$ architectures and trained on $14$ different datasets. For audio source tracing, the MLAAD dataset\footnote{\url{https://deepfake-total.com/sourcetracing}} consists of $11,100$ training, $12,000$ development, and $33,900$ evaluation samples, covering $38$ languages and $82$ TTS models across $33$ different architectures, as detailed in \cite{UsingMLAADforSourceTracing}, totalling $378$ hours of synthetic speech. To mitigate the domain covariate shift between the training and testing domains, we applied offline data augmentation to speech utterances using MUSAN \cite{musan} and Room Impulse Response (RIR) \cite{rir}. This augmentation was performed for original, reverberation, speech, music, and noise. In addition to the MLAAD dataset, we used the ASVSpoof 2019 dataset in the FD and RE training stage \cite{todisco19_interspeech} to enhance the generalization capabilities of systems. 

\subsection{Experimental set-up}
\label{implement}

We used an embedding dimension of $144$ and $24$ classification labels provided in the training set for all the classification networks. For Conformer-based approaches, we utilized $4$ attention heads, a kernel size of $31$, and $4$ Conformer encoders. For state-space models (SSM), including MAMBA and HYDRA, we employed $64$ state spaces and an SSM encoder dimension of $144$. All speech utterances were sampled at $16,000$ Hz and used $4$ seconds of speech segments across train, development and test sets. This study utilizes baseline systems based on the Wav2Vec2-AASIST model, as presented by Xie et al. \cite{baseline_refd}, with an implementation available on GitHub\footnote{\url{https://github.com/piotrkawa/audio-deepfake-source-tracing}}. All systems were trained using a learning rate of $1e-3$, a weight decay of $5e-4$, and a learning rate decay of $0.5$ with the Adam optimizer. The training was conducted over $50$ epochs, with model checkpoints selected based on accuracy performance on the validation set. We used a batch size of $128$ along the gradient accumulation over $8$ batches. The experiments were conducted on a single A100 GPU, with each system requiring approximately $24$ hours of training. All training was conducted using the cross-entropy loss criterion in a multi-class classification setting on the training dataset. We applied a $\beta$ weight of $0$ to the multi-class N-pair loss for $20$ epochs. Afterwards, the weight was initialized to $1e-3$ and updated after every epoch until it reached $0.8$ at epoch $50$.

\subsection{Evaluation schema}
\label{evalschema}
In this work, we evaluate the performance of audio source tracing systems for both in-domain and OOD scenarios using multiple metrics, including Accuracy, F1-score, equal error rate (EER), Negative Log-Likelihood (NLL), Expected Calibration Error (ECE), and Fr\'echet Distance \cite{FD1}. These metrics collectively assess classification performance, uncertainty estimation, calibration, and feature distribution similarity, ensuring a comprehensive evaluation of generalization capabilities. ECE measures the difference between predicted confidence and actual accuracy, offering insight into model calibration. In ECE, predictions are grouped into \( M \) bins, and the discrepancy is computed as:

\begin{equation}
    \text{ECE} = \sum_{m=1}^{M} \frac{|B_m|}{N} \left| \text{acc}(B_m) - \text{conf}(B_m) \right|
\end{equation}

\noindent where \( B_m \) is the set of samples in bin \( m \), and \( \text{acc}(B_m) \) and \( \text{conf}(B_m) \) represent the accuracy and average confidence within that bin respectively.


Fr\'echet Distance has been popularly used in music and video analytics \cite{FD2, FD3}. Fr\'echet Distance measures the similarity between two probability distributions by comparing their statistical properties, specifically their mean and covariance. It quantifies how much the distributions of features from in-domain and OOD data differ, with lower values indicating higher similarity and better model generalization. It is computed as:

\begin{equation} 
    \mathcal{D}_{Frechet} = || \mu_I - \mu_{O} ||^{2} + 
    \text{Tr} \big( \Sigma_I + \Sigma_{O} - 2 (\Sigma_I \Sigma_{O})^{1/2} \big)
\end{equation}

\noindent where \( (\mu_I, \Sigma_I) \) and \( (\mu_{O}, \Sigma_{O}) \) represent the mean and covariance of the in-domain (I), and OOD (O) respective embedding distributions.


\section{Results and analysis}

\begin{table*}[!th]
    \centering
    \caption{Performance evaluation for in-domain and OOD scenarios; Params. refers to model parameters which are in millions(M).}
    \resizebox{\textwidth}{!}{
    \begin{tabular}{clccccccccccc}
        \toprule
        \multirow{2}{*}{\textbf{ID}} & \multirow{2}{*}{\textbf{Systems}} & \multirow{2}{*}{\textbf{Params.}} & \multirow{2}{*}{\textbf{Fr\'echet Dist.}} & \multicolumn{4}{c}{\textbf{Eval set: In-domain}} & \multicolumn{5}{c}{\textbf{Eval set: Out-of-domain}} \\
        \cmidrule(lr){5-8} \cmidrule(lr){9-13}
        & & & & \textbf{Accuracy\%} & \textbf{F1-Score\%} & \textbf{NLL} & \textbf{ECE\%} & \textbf{Accuracy\%} & \textbf{F1-Score\%} & \textbf{EER\%} & \textbf{NLL} & \textbf{ECE\%} \\
        \midrule
        \textbf{} & {Wav2Vec2-AASIST} & 317.8 & 15.05 & 83.39 & 83.34 & -0.7944 & 3.825 & 26.51 & 30.15 & 73.49 & -1.5682 & 0.13 \\
        \midrule
        S1 & XLSR-conformer no RE & 318.1 & 7.36 & 90.74 & 91.12 & -0.8503 & 2.35 & 37.74 & 42.81 & 61.54 & -0.444 & 0.08 \\
        S2 & XLSR-conformer & 318.1 & 8.45 & 95.06 & 95.17 & -0.9255 & 0.56 & 39.54 & 45.15 & 59.21 & -1.3453 & 1.19 \\
        S3 & XLSR-conformer-N-pair & 318.1 & \textbf{6.93} & 95.09 & 95.27 & -0.9243 & \textbf{0.05} & 38.89 & 43.25 & 61.11 & -0.9998 & 0.05 \\
        S4 & XLSR-MAMBA & 317.9 & 49.27 & 79.92 & 77.04 & -0.7041 & 3.68 & 39.29 & 43.66 & 60.11 & \textbf{-7.2452} & 0.06 \\
        S5 & XLSR-HYDRA & 319.7 & 23.04 & 72.01 & 69.77 & -0.5569 & 9.44 & \textbf{44.82} & \textbf{49.31} & \textbf{55.17} & -3.8017 & \textbf{0.01} \\
        \midrule
        E1 & {S2, S3} & --- & 14.72 & \textbf{95.61} & \textbf{95.76} & \textbf{-0.9249} & 1.38 & 38.09 & 42.43 & 61.91 & -0.6039 & 0.06 \\
        E2 & {S3, S4} & --- & 13.86 & 94.25 & 94.22 & -0.8142 & 10.49 & 32.66 & 34.77 & 69.26 & -4.9072 & 0.06 \\
        E3 & {S3, S5} & --- & 13.95 & 94.6 & 94.63 & -0.7406 & 18.73 & 38.04 & 39.92 & 61.34 & -3.138 & \textbf{0.01} \\
        E4 & {S4, S5} & --- & 38.55 & 80.53 & 77.81 & -0.6035 & 12.55 & 39.62 & 44.29 & 60.38 & -5.5271 & 0.06 \\
        \bottomrule
    \end{tabular}}
    \vspace{-1.5em}
\end{table*}

Table 1 presents the performance of the proposed work in both in-domain and OOD scenarios. We adopted the evaluation measures stated in Section \ref{evalschema}. In addition to the proposed approach, we experimented with ensemble fusion systems, denoted by \textit{E}. For the ensemble fusion, we concatenated the respective embeddings from two systems and computed the average logit probability scores across them. The baseline system, Wav2vec2-AASIST \cite{baseline_refd}, and the proposed systems are referred to in Table 1 as follows: XLSR-Conformer without Real Emphasis (RE) (\textit{S1}) represents Stage 1 training, where the model was trained directly using only the Fake Dispersion (FD) loss criterion. XLSR-Conformer trained with Real Emphasis and Fake Dispersion methods as denoted by (\textit{S2}). System \textit{S3} was trained using RE, FD, and multi-class N-pair loss with the XLSR-Conformer model. Systems \textit{S4} and \textit{S5} were trained using only the RE and FD stages. We conducted experiments on the multi-class N-pair loss based on the results obtained from the RE and FD training stages and selected the Conformer-based approach accordingly.

\subsection{In-domain evaluation}

Evaluating system performance within the same data distribution as training provides insights into their effectiveness in controlled settings. The highest-performing model, \textit{E1: S2, S3}, demonstrates superior accuracy of \textit{95.61\%} and an F1-score of \textit{95.76\%}. A negative log-likelihood (NLL) of \textit{-0.9249} reflects the model’s ability to assign high confidence to correct classifications, and an expected calibration error of \textit{1.38\%} indicates a well-calibrated probability estimation. In comparison, the baseline system with an accuracy of \textit{83.39\%} and an F1-score of \textit{83.34\%} performs significantly worse. Individual systems \textit{S2} and \textit{S3} also exhibit superior performance, both achieving over \textit{95\%} accuracy. These results highlight the advantages of model fusion in enhancing deepfake source detection within controlled conditions, ultimately improving classification reliability.

\subsection{Out-of-domain evaluation}
Assessing model generalization to previously unseen deepfake audio is crucial for real-world applicability. The best-performing model in this scenario, \textit{S5}, achieves an accuracy of \textit{44.82\%} and an F1-score of \textit{49.31\%}. Notably, it also records the lowest ECE of \textit{0.01\%}, suggesting precise confidence estimation. The EER of \textit{55.17\%} further supports its capability to distinguish between different audio manipulations. However, the in-domain best-performing model, \textit{E1 (S2, S3)}, suffers a performance drop, achieving only \textit{38.09\%} accuracy and an F1-score of \textit{42.43\%} on OOD data. 
This decline underscores the challenge of balancing in-domain optimization with generalization. The baseline system performs even worse, with an OOD accuracy of merely \textit{26.51\%}, reinforcing the necessity for models that maintain robust performance across diverse deepfake sources.

\subsection{Fr\'echet Distance Analysis}

The Fr\'echet Distance assesses the stability of feature representations, a critical factor in deepfake audio attribution. The model with the lowest Fr\'echet Distance is \textit{S3}, with a value of \textit{6.93}, indicating highly stable and generalizable feature extraction. Although \textit{S3} achieves remarkable in-domain accuracy (\textit{95.09\%}) and F1-score (\textit{95.27\%}), it does not attain the highest OOD accuracy, registering \textit{38.89\%}. However, its consistent feature representations across all the domains position it as a viable candidate for source tracing tasks. Understanding the correlation among NLL, ECE, accuracy, and F1-score is vital in selecting an optimal model. A lower NLL implies well-calibrated probability predictions, while a lower ECE signifies reliable confidence estimation. Accuracy and F1-score directly reflect classification performance, with F1-score being particularly important in deepfake detection due to its balance between precision and recall, minimizing the false positives and false negatives. For applications emphasizing in-domain accuracy, \textit{E1 (S2, S3)} is the most suitable choice. If generalization to OOD data is prioritized, \textit{S5} stands out. However, considering feature stability and classification effectiveness, \textit{S3} emerges as the most balanced model. These findings indicate the ongoing challenge of developing models that perform reliably in both in-domain and OOD settings for deepfake audio source tracing.

\section{Discussion}

The evaluation results highlight the system's performance across different conditions. The proposed methods showed significant improvements over the baseline in both in-domain and OOD scenarios. Among the models proposed, the Conformer model outperformed state-space models like MAMBA and HYDRA. Ablation experiments with three training strategies showed that \textit{S3}, achieved the best generalization, with optimal Fr\'echet Distance and ECE, underscoring the importance of meaningful synthetic audio representations for OOD robustness. Ensemble fusion (E1), which combined logit scores and embeddings, excelled in in-domain tasks while maintaining OOD performance close to \textit{S3}. A trade-off between in-domain and OOD performance was observed, particularly in the HYDRA-based system (S5), which performed best in OOD scenarios, suggesting the need for regularization techniques in future. 

Fr\'echet Distance was validated as a reliable single metric for evaluating embedding distribution differences, correlating lower Fr\'echet Distance with better system performance. Future research should explore a broader range of audio deepfake sources and utilize anti-spoofing datasets to enhance robustness. The study demonstrated the effectiveness of deep metric learning, emphasizing the potential of contrastive learning frameworks to further improve generalization and create a more structured latent space for audio deepfake embeddings.

\section{Conclusion}

In this paper, we presented a two-stage training approach, utilizing Real Emphasis and Fake Dispersion strategies, coupled with the novel multi-class N-pair loss, ensuring robust generalization across different deepfake sources. Our experiments demonstrated that while the best-performing models (S2, S3) exhibit strong in-domain performance, challenges persist in generalizing to out-of-domain (OOD) data. Despite the accuracy drop in OOD evaluation, the S3 model stands out with its stable feature representations, as indicated by the low Fr\'echet Distance, making it a promising candidate for source tracing tasks that require consistency across domains. The fusion models emphasize the importance of ensemble strategies, enhancing the trade-off between in-domain and out-of-domain source tracing scenarios. Overall, this work highlights the trade-offs between in-domain accuracy and OOD generalization, underscoring the importance of feature stability and calibration in real-world applications. 

Future research will focus on regularization techniques and incorporating additional audio sources from existing anti-spoofing datasets to enhance generalization to OOD scenarios while maintaining high performance in controlled settings. This will enhance the meaningful latent space representation for audio source tracing techniques.


\pagebreak

\section{Acknowledgements}
This work was partially supported by the Innosuisse through the flagship project IICT: Inclusive Information and Communication Technologies (grant agreement no. PFFS-21-47), by the Swiss National Science Foundation through the project PASS: Pathological Speech Synthesis (grant agreement no. 219726), and by the Estonian Centre of Excellence in Artificial Intelligence (EXAI).


\vspace{-0.5em}


\bibliographystyle{IEEEtran}
\bibliography{mybib}

\begin{thebibliography}{10}
\providecommand{\url}[1]{#1}
\csname url@samestyle\endcsname
\providecommand{\newblock}{\relax}
\providecommand{\bibinfo}[2]{#2}
\providecommand{\BIBentrySTDinterwordspacing}{\spaceskip=0pt\relax}
\providecommand{\BIBentryALTinterwordstretchfactor}{4}
\providecommand{\BIBentryALTinterwordspacing}{\spaceskip=\fontdimen2\font plus
\BIBentryALTinterwordstretchfactor\fontdimen3\font minus \fontdimen4\font\relax}
\providecommand{\BIBforeignlanguage}[2]{{%
\expandafter\ifx\csname l@#1\endcsname\relax
\typeout{** WARNING: IEEEtran.bst: No hyphenation pattern has been}%
\typeout{** loaded for the language `#1'. Using the pattern for}%
\typeout{** the default language instead.}%
\else
\language=\csname l@#1\endcsname
\fi
#2}}
\providecommand{\BIBdecl}{\relax}
\BIBdecl

\bibitem{knntts}
\BIBentryALTinterwordspacing
K.~E. Hajal, A.~Kulkarni, E.~Hermann, and M.~Magimai~Doss, ``k{NN} retrieval for simple and effective zero-shot multi-speaker text-to-speech,'' in \emph{NAACL}, L.~Chiruzzo, A.~Ritter, and L.~Wang, Eds., 2025. [Online]. Available: \url{https://aclanthology.org/2025.naacl-short.65/}
\BIBentrySTDinterwordspacing

\bibitem{knnvcr}
K.~E. Hajal, A.~Kulkarni, E.~Hermann, and M.~M. Doss, ``Unsupervised rhythm and voice conversion of dysarthric to healthy speech for asr,'' in \emph{Workshop on Speech Pathology Analysis and DEtection (SPADE), IEEE}, 2025.

\bibitem{todisco19_interspeech}
M.~Todisco, X.~Wang, V.~Vestman, M.~Sahidullah, H.~Delgado, A.~Nautsch, J.~Yamagishi, N.~Evans, T.~H. Kinnunen, and K.~A. Lee, ``Asvspoof 2019: Future horizons in spoofed and fake audio detection,'' in \emph{proc. of Interspeech}, 2019.

\bibitem{kulkarni24_asvspoof}
A.~Kulkarni, H.~M. Tran, A.~Kulkarni, S.~Dowerah, D.~Lolive, and M.~M. Doss, ``Exploring generalization to unseen audio data for spoofing: insights from ssl models,'' in \emph{The Automatic Speaker Verification Spoofing Countermeasures Workshop (ASVspoof 2024)}, 2024.

\bibitem{Wang2019ASVspoof2A}
X.~Wang, J.~Yamagishi, M.~Todisco, H.~Delgado, A.~Nautsch, N.~W.~D. Evans, M.~Sahidullah, V.~Vestman, T.~H. Kinnunen, K.~A. LEE, L.~Juvela, P.~Alku, Y.-H. Peng, H.-T. Hwang, Y.~Tsao, H.-M. Wang, S.~L. Maguer, M.~Becker, and Z.~Ling, ``Asvspoof 2019: A large-scale public database of synthesized, converted and replayed speech,'' \emph{Computer Speech \& Language}, vol.~64, 2019.

\bibitem{Borrelli2021SyntheticSD}
C.~Borrelli, P.~Bestagini, F.~Antonacci, A.~Sarti, and S.~Tubaro, ``Synthetic speech detection through short-term and long-term prediction traces,'' \emph{proc. of EURASIP Journal on Information Security}, 2021.

\bibitem{zhang2024distinguishingneuralspeechsynthesis}
C.~Y. Zhang, J.~Yi, J.~Tao, C.~Wang, and X.~Yan, ``Distinguishing neural speech synthesis models through fingerprints in speech waveforms,'' \emph{Proceedings of the 23rd Chinese National Conference on Computational Linguistics}, 2024.

\bibitem{zhu2022sourcetracingdetectingvoice}
T.~Zhu, X.~Wang, X.~Qin, and M.~Li, ``Source tracing: Detecting voice spoofing,'' \emph{Asia-Pacific Signal and Information Processing Association Annual Summit and Conference}, 2022.

\bibitem{Klein2024SourceTO}
N.~Klein, T.~Chen, H.~Tak, R.~Casal, and E.~Khoury, ``Source tracing of audio deepfake systems,'' \emph{proc. of Interspeech}, 2024.

\bibitem{Yi2023ADD2T}
J.~Yi, J.~Tao, R.~Fu, X.~Yan, C.~Wang, T.~Wang, C.~Y. Zhang, X.~Zhang, Y.~Zhao, Y.~Ren, and L.~Xu, ``Add 2023: the second audio deepfake detection challenge,'' in \emph{proc. of DADA@IJCAI}, 2023.

\bibitem{nast}
\BIBentryALTinterwordspacing
Y.~Xie, X.~Wang, Z.~Wang, R.~Fu, Z.~Wen, S.~Cao, L.~Ma, C.~Li, H.~Cheng, and L.~Ye, ``Neural codec source tracing: Toward comprehensive attribution in open-set condition,'' vol. abs/2501.06514, 2025. [Online]. Available: \url{https://arxiv.org/abs/2501.06514}
\BIBentrySTDinterwordspacing

\bibitem{baseline_refd}
Y.~Xie, R.~Fu, Z.~Wen, Z.~Wang, X.~Wang, H.~Cheng, L.~Ye, and J.~Tao, ``Generalized source tracing: Detecting novel audio deepfake algorithm with real emphasis and fake dispersion strategy,'' in \emph{proc. of Interspeech}, 2024.

\bibitem{FD1}
M.~PoceviVCiūtė, G.~Eilertsen, S.~Garvin, and C.~Lundstrom, ``Detecting domain shift in multiple instance learning for digital pathology using fréchet domain distance,'' in \emph{International Conference on Medical Image Computing and Computer-Assisted Intervention}, 2024.

\bibitem{FD2}
J.~Liu, Y.~Qu, Q.~Yan, X.~Zeng, L.~Wang, and R.~Liao, ``Fréchet video motion distance: A metric for evaluating motion consistency in videos,'' \emph{ArXiv}, vol. abs/2407.16124, 2024.

\bibitem{conformer}
A.~Gulati, J.~Qin, C.-C. Chiu, N.~Parmar, Y.~Zhang, J.~Yu, W.~Han, S.~Wang, Z.~Zhang, Y.~Wu, and R.~Pang, ``Conformer: Convolution-augmented transformer for speech recognition,'' in \emph{proc. of Interspeech}, 2020.

\bibitem{npair}
K.~Sohn, ``Improved deep metric learning with multi-class n-pair loss objective,'' in \emph{proc. of Neural Information Processing Systems}, 2016.

\bibitem{mamba}
\BIBentryALTinterwordspacing
A.~Gu and T.~Dao, ``Mamba: Linear-time sequence modeling with selective state spaces,'' in \emph{First Conference on Language Modeling}, 2024. [Online]. Available: \url{https://openreview.net/forum?id=tEYskw1VY2}
\BIBentrySTDinterwordspacing

\bibitem{hydra}
\BIBentryALTinterwordspacing
S.~Hwang, A.~Lahoti, R.~Puduppully, T.~Dao, and A.~Gu, ``Hydra: Bidirectional state space models through generalized matrix mixers,'' in \emph{proc. of Neural Information Processing Systems}, 2024. [Online]. Available: \url{https://openreview.net/forum?id=preo49P1VY}
\BIBentrySTDinterwordspacing

\bibitem{Baevski2020wav2vec2A}
A.~Baevski, H.~Zhou, A.~rahman Mohamed, and M.~Auli, ``wav2vec 2.0: A framework for self-supervised learning of speech representations,'' \emph{proc. of Neural Information Processing Systems}, 2020.

\bibitem{tcm}
D.-T. Truong, R.~Tao, T.~Nguyen, H.-T. Luong, K.~A. Lee, and E.~S. Chng, ``Temporal-channel modeling in multi-head self-attention for synthetic speech detection,'' in \emph{proc. of Interspeech}, 2024.

\bibitem{sls}
Q.~Zhang, S.~Wen, and T.~Hu, ``Audio deepfake detection with self-supervised xls-r and sls classifier,'' \emph{Proceedings of the 32nd ACM International Conference on Multimedia}, 2024.

\bibitem{mambadf}
Y.~Xiao and R.~K. Das, ``Xlsr-mamba: A dual-column bidirectional state space model for spoofing attack detection,'' \emph{ArXiv}, vol. abs/2411.10027, 2024.

\bibitem{conformersv}
Y.~Zhang, Z.~Lv, H.~Wu, S.~Zhang, P.~Hu, Z.~Wu, H.~yi~Lee, and H.~Meng, ``Mfa-conformer: Multi-scale feature aggregation conformer for automatic speaker verification,'' in \emph{proc. of Interspeech}, 2022.

\bibitem{transformer}
A.~Vaswani, N.~M. Shazeer, N.~Parmar, J.~Uszkoreit, L.~Jones, A.~N. Gomez, L.~Kaiser, and I.~Polosukhin, ``Attention is all you need,'' in \emph{proc. of Neural Information Processing Systems}, 2017.

\bibitem{ocsoftmax}
Y.~Zhang, F.~Jiang, and Z.~Duan, ``One-class learning towards synthetic voice spoofing detection,'' \emph{IEEE Signal Processing Letters}, 2020.

\bibitem{vrm}
O.~Chapelle, J.~Weston, L.~Bottou, and V.~Vapnik, ``Vicinal risk minimization,'' in \emph{proc. of Neural Information Processing Systems}, T.~Leen, T.~Dietterich, and V.~Tresp, Eds., 2000.

\bibitem{erm}
V.~N. Vapnik, ``Principles of risk minimization for learning theory,'' in \emph{proc. of Neural Information Processing Systems}, 1991.

\bibitem{regmix}
F.~Pinto, H.~Yang, S.~N. Lim, P.~H.~S. Torr, and P.~K. Dokania, ``Using mixup as a regularizer can surprisingly improve accuracy \& out-of-distribution robustness,'' in \emph{proc. of Neural Information Processing Systems}, 2022.

\bibitem{metriclearning}
M.~Kaya and H.~Şakir Bilge, ``Deep metric learning: A survey,'' \emph{proc. of Symmetry}, vol.~11, p. 1066, 2019.

\bibitem{mlaad}
N.~M. M{\"u}ller, P.~Kawa, W.~H. Choong, E.~Casanova, E.~G{\"o}lge, T.~M{\"u}ller, P.~Syga, P.~Sperl, and K.~B{\"o}ttinger, ``Mlaad: The multi-language audio anti-spoofing dataset,'' \emph{International Joint Conference on Neural Networks (IJCNN)}, 2024.

\bibitem{UsingMLAADforSourceTracing}
N.~M{\"u}ller, ``Using mlaad for source tracing of audio deepfakes,'' \url{https://deepfake-total.com/sourcetracing}, Fraunhofer AISEC, 11 2024.

\bibitem{musan}
D.~Snyder, G.~Chen, and D.~Povey, ``Musan: A music, speech, and noise corpus,'' \emph{ArXiv}, vol. abs/1510.08484, 2015.

\bibitem{rir}
T.~Ko, V.~Peddinti, D.~Povey, M.~L. Seltzer, and S.~Khudanpur, ``A study on data augmentation of reverberant speech for robust speech recognition,'' \emph{proc. of ICASSP}, 2017.

\bibitem{FD3}
J.~Retkowski, J.~Stepniak, and M.~Modrzejewski, ``Frechet music distance: A metric for generative symbolic music evaluation,'' \emph{ArXiv}, vol. abs/2412.07948, 2024.

\end{thebibliography}

\end{document}